\documentclass{jetpl}

\usepackage{graphicx}

\twocolumn \lat

 \DeclareMathOperator{\sgn}{sgn}
 \DeclareMathOperator{\Tr}{Tr}

 \title{Conductance of a junction between a normal metal\\ and a Berezinskii superconductor}
 \rtitle{Conductance of a junction between a normal metal and a Berezinskii superconductor}
 \sodtitle{Conductance of a junction between a normal metal and a Berezinskii superconductor}

 \author{Ya.\,V.~Fominov\/\thanks{e-mail: fominov@landau.ac.ru}}
 \rauthor{Ya.\,V.~Fominov}
 \sodauthor{Fominov}

\address{L.\,D.~Landau Institute for Theoretical Physics RAS, 119334 Moscow, Russia}

\dates{15 March 2008}{*}

\abstract{The conductance of a junction between a normal metal and a superconductor having the symmetry proposed by
Berezinskii is studied theoretically. The main feature of this symmetry is the odd frequency dependence of the anomalous
Green function, which makes possible the \textit{s}-wave triplet superconducting state (the Berezinskii superconductor).
The Andreev reflection (which links positive and negative energies) is sensitive to the energetic symmetry; as a result,
the conductance of the junction involving the Berezinskii superconductor is qualitatively different from the case of a
conventional superconductor. Experimentally, the obtained results can be employed to test the possibility of the
Berezinskii superconductivity proposed for Na$_x$CoO$_2$ and to identify the odd-$\omega$ component predicted for
superconductor-ferromagnet systems.\\ \centerline{\textbf{Published as: JETP Letters \textbf{86}, 732 (2007) [Pis'ma v
ZhETF \textbf{86}, 842 (2007)]}}}


\begin{document}
\maketitle

The superconducting pairing can be described by the anomalous Green function $F(1;2)$ (also referred to as the pair wave
function) in the Matsubara technique. The Pauli principle requires antisymmetry under permutation of two electrons in a
Cooper pair, $F(2;1) = -F(1;2)$, which leads to the standard classification of superconducting phases \cite{Mineev}: if
the coordinate dependence of $F$ is even (\textit{s}-wave, \textit{d}-wave, etc.), then the spin dependence must be odd
(singlet) and vice versa. This assumes that $F$ is an even function of the imaginary time $\tau = \tau_1-\tau_2$ or of
the frequency $\omega$ in the Fourier representation.

In 1974, Berezinskii suggested \cite{Berez} that the \textit{s}-wave triplet phase (not listed above) is also possible
if $F$ is an odd function of $\omega$. The Pauli principle is satisfied and there are no symmetry restrictions on the
existence of such a phase. Of course, the question remains whether an interaction necessary for the realization of the
odd-$\omega$ phase exists in real materials. Berezinskii discussed the odd-$\omega$ phase as a possible phase of $^3$He
and argued (without a full microscopic derivation) that it could be formed due to the retarded paramagnon exchange.
However, up to now there are no indications for such a state in $^3$He. Moreover, there is no microscopic theory
producing the Berezinskii state in a bulk material. At the same time, properties of hypothetic odd-$\omega$
superconductors were studied in a number of papers, e.g. \cite{AB}.

Surprisingly, in 2001 the ``long-range'' Berezinskii superconductivity was theoretically discovered by Bergeret, Volkov,
and Efetov \cite{BVElong} (see also \cite{Kadigrobov}) in a SF system consisting of a conventional (\textit{s}-wave
singlet) superconductor and a ferromagnet. They demonstrated that the superconducting component with the symmetry
proposed by Berezinskii arises in the case of inhomogeneous magnetization of the ferromagnet due to the proximity effect
and penetrates the ferromagnet over a much longer distance compared to the singlet component, since the exchange field
does not suppress the triplet superconducting correlations with projections $S_z = \pm 1$. This feature enables to
spatially single out the Berezinskii component. Many unusual properties of this behavior have been recently investigated
\cite{BVE_review}. At the same time, experimental verification of the $S_z=\pm 1$ component in SF structures is still
under debate. A number of experiments observed a long-range proximity effect \cite{long-range} and a Josephson coupling
through a half-metallic ferromagnet \cite{half-metal}, which can be interpreted in terms of the long-range component.
However, the data on the structure of the magnetic inhomogeneity in experimental samples is lacking.

At the same time, as argued in \cite{Mazin}, the Berezinskii superconductivity can be realized in Na$_x$CoO$_2$. This
conjecture is based on the band-structure calculations and on the experimental results which indicate the triplet
superconductivity (from the Knight-shift data) and the \textit{s}-wave symmetry (from insensitivity to impurities).

The question arises: can we suggest an experiment which is sensitive to the most nontrivial feature of the Berezinskii
superconductivity, the odd-$\omega$ dependence? If yes, then this experiment could be a good test for such a state,
similarly to the famous experiments sensitive to the nontrivial spatial symmetry of anisotropic superconductivity (the
idea proposed in \cite{Geshkenbein} was employed to experimentally verify the \textit{d}-wave symmetry in YBaCuO, see
review \cite{VanHarlingen}). It seems natural that the odd-$\omega$ superconductivity should lead to unusual features of
the Andreev reflection \cite{Andreev} from such a superconductor, since this process links an electron with positive
energy to a hole with negative energy (with respect to the Fermi level).

\begin{figure}
 \centerline{\includegraphics[width=4cm]{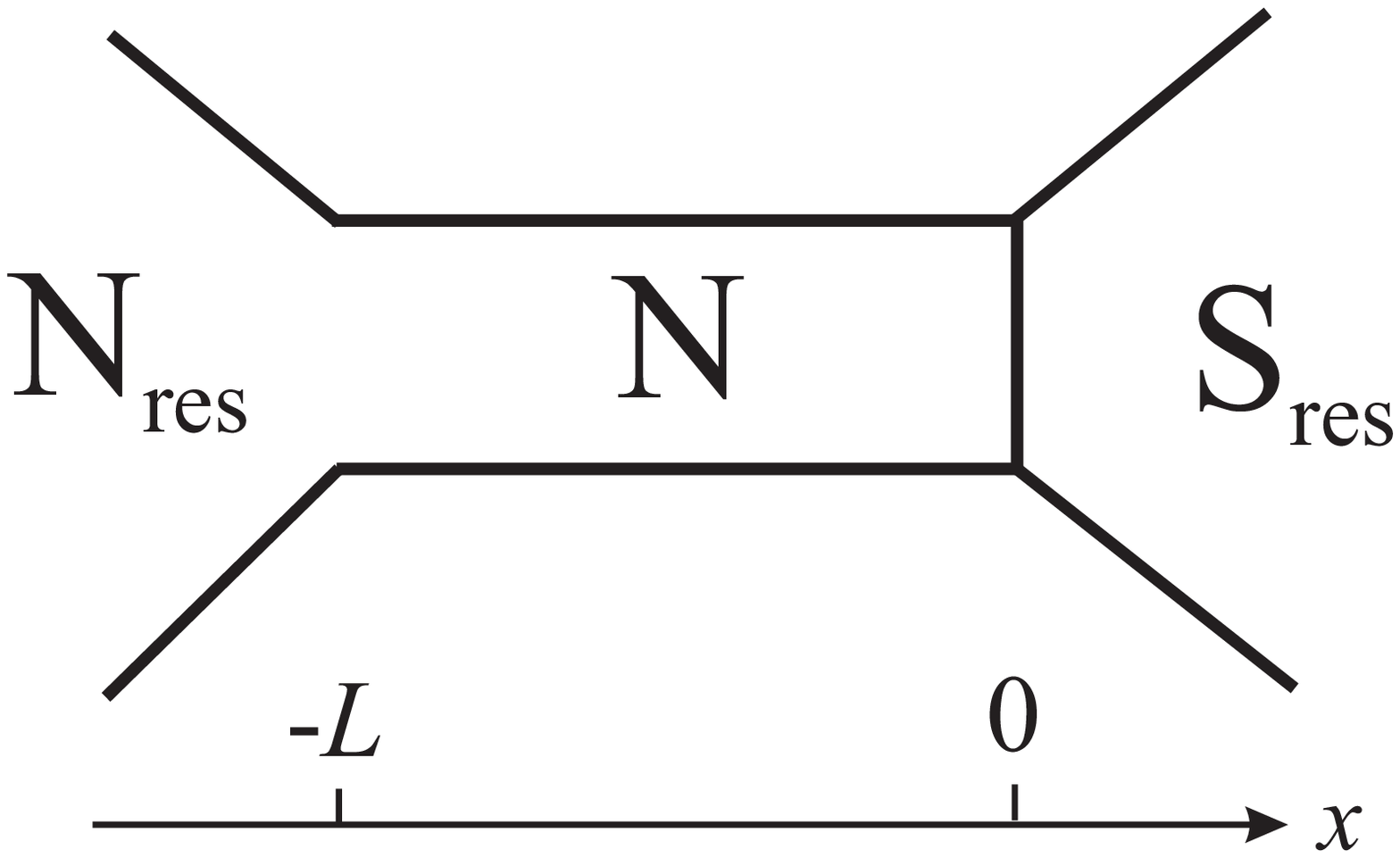}}
\caption{Fig.\ref{fig:junction}. NS junction. The conductances of the normal wire and the NS interface are $G_N$ and
$G_T$, respectively. The voltage $V$ is applied to the normal reservoir N$_\mathrm{res}$.}
 \label{fig:junction}
\end{figure}

In this Letter, the differential conductance of the normal-metal---superconductor (NS) junction shown in
Fig.~\ref{fig:junction} is studied at zero temperature. I consider three possibilities for the superconducting
reservoir: S~--- a conventional superconductor with a gap (in this case previous results are reproduced, see reviews
\cite{LR-PC} and references therein), S$_0$~--- a conventional superconductor without a gap (e.g., due to paramagnetic
impurities \cite{AG}), and S$_B$~--- a Berezinskii superconductor. S$_B$ has two main features: odd-$\omega$ symmetry
and gapless spectrum. The case of S$_0$ is considered in order to reveal the features of S$_B$ which are due to the
unusual symmetry and not simply due to its gapless spectrum. Physical examples of the S$_B$ state are the bulk
Berezinskii superconductor and the SF system (in the latter case, the normal wire should be attached to a region where
only the long-range triplet component survives, while the short-range ones are negligible).

The differential conductance of the junction is
\begin{equation}
G_{NS} (V) = \frac{dI(V)}{dV} ;
\end{equation}
its normal-state value is $G_0 = \left( G_N^{-1} + G_T^{-1} \right)^{-1}$. The
diffusive normal wire of length $L$ is characterized by the Thouless energy $E_\mathrm{Th} = D/L^2$, where $D$ is the
diffusion constant. At low voltages, the Andreev reflection plays an important role in the transport. I assume $eV$,
$E_\mathrm{Th} \ll E_0$, where $E_0$ is the energy scale on which the Green function of the superconductor varies (in a
conventional superconductor $E_0$ coincides with the static order parameter $\Delta$). At the same time, the relation
between $eV$ and $E_\mathrm{Th}$ is arbitrary.

I consider the dirty limit and employ the Usadel equation \cite{Usadel,Kopnin} for the Green function $g$ which is a
$8\times 8$ matrix in the Keldysh $\otimes$ Nambu-Gor'kov $\otimes$ spin space:
\begin{equation}
g = \begin{pmatrix} \check g^R & \check g^K \\ 0 & \check g^A \end{pmatrix} .
\end{equation}
The Usadel equation for $g$ is
\begin{equation}
D \nabla ( g \nabla g) + \left[ i E \hat\tau_3 \hat\sigma_0 \hat 1, g \right] =0,
\end{equation}
where $\hat\tau$ and $\hat\sigma$ denote the Pauli matrices in the Nambu-Gor'kov and spin spaces, respectively, while
$\hat 1$ is the unit matrix in the Keldysh space. At the interface with the normal reservoir ($x=-L$), $g$ must be equal
to the bulk normal-metallic function, while the Kupriyanov-Lukichev boundary condition \cite{Kupriyanov} at the NS
interface ($x=0$) reads
\begin{equation} \label{KL}
g \nabla g = \frac{G_T}{2 G_N L} \left[ g, g_S \right],
\end{equation}
where $g_S$ is the Green function in the superconductor.

The current is $I = \int j dE$ with the spectral current
\begin{equation} \label{j}
j(E,x) = - \frac{G_N L}{16e} \Tr\left[ \hat\tau_3 \hat\sigma_0 ( g \nabla g )^K \right].
\end{equation}

Due to the normalization condition $g^2 =1$, we can parametrize $\check g^K$ as $\check g^K = \check g^R \check f -
\check f \check g^A$, then the general relation $\check g^A = - \hat\tau_3 \left( \check g^R \right)^\dagger \hat\tau_3$
allows us to consider only $\check g^R$ and $\check f$ as independent functions.

In S and S$_0$, we can write
\begin{equation} \label{g_S}
\check g^R = \hat\tau_3 \hat\sigma_0 G^R + \hat\tau_1 \hat\sigma_0 F^R,
\end{equation}
while S$_B$ is a spin-triplet superconductor and the anomalous part is a vector in the spin space. Choosing its
direction as $z$, we write \cite{z}
\begin{equation} \label{g_SB}
\check g^R = \hat\tau_3 \hat\sigma_0 G^R + \hat\tau_1 \hat\sigma_3 F^R.
\end{equation}
Physically, this is a triplet superconducting state with zero projection of the Cooper pair's spin on the $z$ axis,
while the $1$ and $-1$ projections on any axis in the $xy$ plane are equiprobable.

The Berezinskii superconductivity is odd in the Matsubara frequency, $F(-\omega) = -F(\omega)$, therefore in the
real-energy representation we obtain $F^R(-E) = - F^A (E)$. Together with the general relation $F^A (E) = \left( F^R (E)
\right)^*$ this yields \cite{TG} $F^R(-E) = -\left( F^R (E) \right)^*$. We need to know the low-energy behavior of
$F^R(E)$, which depends on the choice of a model. The question of a microscopic model for a bulk Berezinskii
superconductor is not clear at present. At the same time, we know that such a state is realized in inhomogeneous SF
structures due to the proximity effect. Taking the low-energy behavior of the odd-$\omega$ triplet component from, e.g.,
\cite{FVE}, we obtain $F(\omega) = iA \sgn\omega$, hence $F^R(E) = iA$ at $E\to 0$, with real constant $A$. This
behavior also follows from the standard relation $F^R = \Delta/\sqrt{\Delta^2-E^2}$ in the case of the linear low-energy
behavior $\Delta (E) = E / (1+A^{-2})$.

The constraint $(\check g^R)^2 =1$ allows us to parametrize the normal and anomalous Green functions by a (complex)
angle $\theta$ as $G^R = \cos\theta$ and $F^R = \sin\theta$ in the cases of both the conventional and Berezinskii
superconductor, Eqs.\ (\ref{g_S}) and (\ref{g_SB}). The equation and the boundary conditions for $\theta$ are
\begin{gather}
\frac D2 \theta'' + i E \sin\theta =0, \label{Usadel} \\
\theta = 0 \Bigr|_{x=-L},\quad \theta' = \frac{G_T}{G_N L} \sin (\theta_S -\theta) \Bigr|_{x=0}. \label{bc0}
\end{gather}
The superconductor is described by $\theta_S$. In order to consider low voltages, we need to find the low-energy
solution of the Usadel equation. If $E \ll E_0$, then $\theta_S = \pi/2$ in~S, $0< \theta_S <\pi/2$ in~S$_0$, and
$\theta_S = i \vartheta_S$ with real $\vartheta_S$ in~S$_B$. Thus the type of the superconductor enters our
consideration only through the low-energy value $\theta_S$.

In the case of conventional superconductors, S or S$_0$, components of the distribution function can be chosen as
\begin{equation}
\check f = \hat\tau_0 \hat\sigma_0 f_0 +  \hat\tau_3 \hat\sigma_0 f_3,
\end{equation}
while in the case of S$_B$ the number of components is doubled in the general case:
\begin{equation}
\check f = \hat\tau_0 \hat\sigma_0 f_0 +  \hat\tau_3 \hat\sigma_0 f_3 + \hat\tau_0 \hat\sigma_3 \bar f_0 + \hat\tau_3
\hat\sigma_3 \bar f_3.
\end{equation}

The spectral current (\ref{j}) at $x=0$ can be rewritten with the help of the boundary condition (\ref{KL}) as
\begin{multline} \label{I}
j(E) = (G_T / 8 e) f_3 \times \\
\times \left[ (G^R-G^A) (G_S^R-G_S^A) + (F^R+F^A) (F_S^R+F_S^A) \right].
\end{multline}
The two terms in the square brackets have clear physical meaning. Since $(G^R-G^A)/2$ is the single-particle density of
states (DOS), the first term in the square brackets describes the quasiparticle contribution to the current. At the same
time, the second term (of $FF$ type) describes the supercurrent due to the Andreev reflection.

The subsequent derivation is similar to the conventional case \cite{VZK,LR-PC}. At zero temperature the integral over
energies, which determines $G_{NS}(V)$, reduces to the sum of the two terms with $E=\pm eV$. Finally, using the symmetry
$\theta (-E) = \theta^*(E)$ for the conventional even-$\omega$ superconductor or $\theta(-E) = - \theta^*(E)$ for the
odd-$\omega$ one \cite{TG}, we obtain
\begin{multline} \label{GNS}
\frac 1{G_{NS} (V)} = \frac 1{G_N L} \int_{-L}^0 \frac{dx}{\cosh^2 \theta_2 (x)} + \\
+ \frac 1{G_T \cos(\theta_{S1}-\theta_1) \cosh\theta_{S2} \cosh\theta_2} ,
\end{multline}
where the right-hand side (r.h.s.) is taken at $E=eV$ and we have separated $\theta$ into the real and imaginary parts,
$\theta = \theta_1 +i\theta_2$. The angles $\theta$ and $\theta_S$ in the second term are taken at the NS interface. The
problem now reduces to solving Eqs.\ (\ref{Usadel})-(\ref{bc0}) and calculating the r.h.s. of Eq.\ (\ref{GNS}).

We start from the simplest case of zero bias $V=0$. At $E=0$, Eqs.\ (\ref{Usadel}) and (\ref{bc0}) are solved by a
linear function. In the case of S or S$_0$ we obtain
\begin{equation}
G_{NS}^{-1}(0) = G_N^{-1} + \frac{G_T^{-1}}{\cos(\theta_S-\theta_0)},
\end{equation}
where $\theta_0$ must be determined from the equation $\theta_0 = (G_T/G_N) \sin(\theta_S-\theta_0)$. In the case of
S$_B$:
\begin{equation}
G_{NS}^{-1}(0) = \frac{G_N^{-1} \tanh\vartheta_0}{\vartheta_0} + \frac{G_T^{-1}}{\cosh\vartheta_0 \cosh\vartheta_S} ,
\end{equation}
where $\vartheta_0$ must be determined from the equation $\vartheta_0 = (G_T/G_N) \sinh (\vartheta_S-\vartheta_0)$.

An immediate consequence of these results is that $G_{NS}(0) < G_0$ in the cases of S and S$_0$, while $G_{NS}(0) > G_0$
for S$_B$. In the S case, where the low-energy DOS is zero, the current is entirely due to the Andreev contribution. In
the S$_0$ case, a finite DOS appears and the current has both the Andreev and quasiparticle contributions.
Interestingly, in the S$_B$ case, only the quasiparticle processes contribute to $G_{NS}(0)$; this fact can be
interpreted as the absence of the Andreev reflection from S$_B$ at $E\to 0$.

At $eV \ll E_\mathrm{Th}$, we can develop a perturbation theory finding corrections to the zero-bias solution of the
Usadel equation. Two orders in $eV / E_\mathrm{Th}$ yield a quadratic low-bias behavior: $G_{NS}(V) = G_{NS}(0) + a
V^2$. The explicit form of $a$ is cumbersome and I only present its most important features. The sign of $a$ depends on
the ratio $G_N / G_T$ and on the type of the superconductor: $a > 0$ for S and S$_0$ if $G_N / G_T < g_c$ and $a < 0$ if
$G_N / G_T > g_c$ where $g_c$ is of the order of unity and weakly depends on $\theta_S$ (as $\theta_S$ changes from
$\pi/2$ to $0$, the critical value $g_c$ stays between $0.8$ and $0.9$), while $a<0$ for S$_B$ at arbitrary $G_N/G_T$.

Now let us consider in more detail the limiting cases of large and small ratio $G_N / G_T$.

\textbf{Tunneling limit:} $G_N \gg G_T$. In this case, we can retain only the second term in the r.h.s.\ of Eq.\
(\ref{GNS}). The proximity effect is weak, $|\theta| \ll 1$.

In the case of S$_0$ and S$_B$ we can set $\theta=0$, then
\begin{equation} \label{ts}
\frac{G_{NS} (V)}{G_0} = \nu_S(eV) \approx \nu_S(0) ,
\end{equation}
where $\nu_S (E)$ is the DOS in the superconductor, normalized to the normal-metallic value; this is nearly constant at
$E\ll E_0$ (the more accurate analysis presented above gives the small bias-dependent correction to this constant). The
physical meaning of Eq.\ (\ref{ts}) is the tunneling spectroscopy of the superconductor with the normal probe. An
essential difference between S$_0$ and S$_B$ is that $\nu_S(0) = \cos\theta_S < 1$ for S$_0$, while $\nu_S(0) =
\cosh\vartheta_S > 1$ for S$_B$ (this fact for S$_B$ was pointed out in \cite{TG}).

In the case of S, $\nu_S(eV)=0$ below the gap, therefore we cannot neglect the proximity effect. Linearizing Eqs.\
(\ref{Usadel}) and (\ref{bc0}) over $\theta$, we find the solution and finally obtain
\begin{equation} \label{G_tun}
\frac{G_{NS} (V)}{G_0} = \frac{G_T}{G_N} \frac{\sinh (2\sqrt\varepsilon) + \sin (2\sqrt\varepsilon)}{4\sqrt\varepsilon
[\sinh^2 (\sqrt\varepsilon) + \cos^2 (\sqrt\varepsilon)]},\quad \varepsilon = \frac{eV}{E_\mathrm{Th}} ,
\end{equation}
hence $G_{NS} (V) \ll G_0$ [Eq.\ (\ref{G_tun}) also follows from a more general result obtained in \cite{VZK}].

\begin{figure}
 \centerline{\includegraphics[width=80mm]{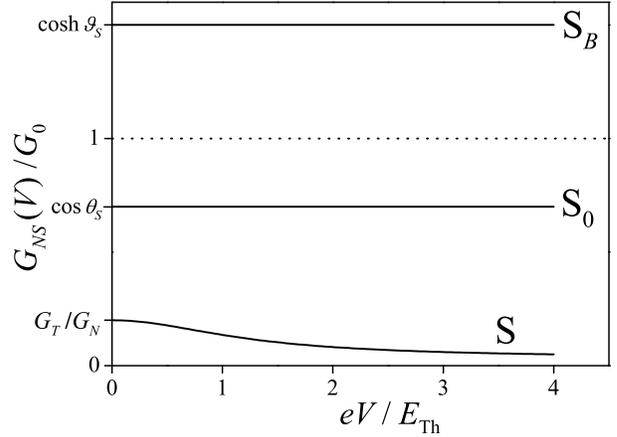}}
\caption{Fig.\ref{fig:G_NS_V_tun}. Differential conductance at $eV\ll E_0$ in the tunneling limit ($G_N \gg G_T$). In
the S case, $G_{NS}(V)$ demonstrates the zero-bias anomaly \cite{ZBA}: it has quadratic and $1/\sqrt V$ dependences at
$eV \ll E_\mathrm{Th}$ and $eV \gg E_\mathrm{Th}$, respectively. In the S$_0$ and S$_B$ cases, $G_{NS}(V)$ is nearly
constant, smaller than $G_0$ for S$_0$ and larger than $G_0$ for S$_B$.}
 \label{fig:G_NS_V_tun}
\end{figure}

The results for the tunneling limit are summarized in Fig.~\ref{fig:G_NS_V_tun}. Note that the physics related to the
Andreev reflection is not essential for S$_B$ in this limit, since the transport is due to the quasiparticle
contribution. At the same time, the Andreev reflection plays a crucial role in the transparent limit considered below.

\begin{figure}
 \centerline{\includegraphics[width=80mm]{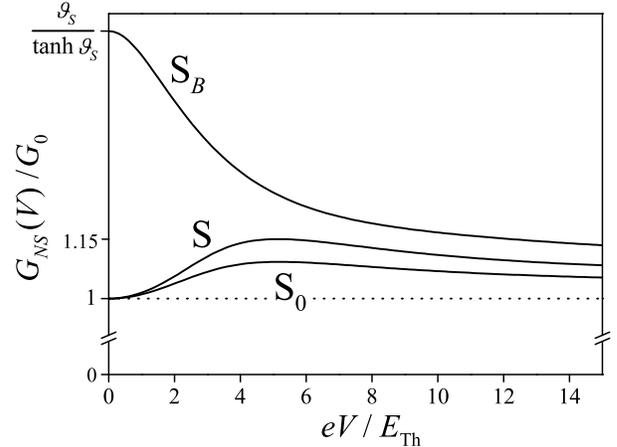}}
\caption{Fig.\ref{fig:G_NS_V}. Differential conductance at $eV\ll E_0$ in the transparent limit ($G_N \ll G_T$). In all
the three cases, $G_{NS}(V)$ is quadratic at $eV \ll E_\mathrm{Th}$ and approaches unity as $1/\sqrt V$ at $eV \gg
E_\mathrm{Th}$. In the S and S$_0$ cases, the behavior is reentrant (see \cite{reentrance} for the S case). The maximum
value of $G_{NS}(V)/G_0$ (achieved at $eV$ of the order of several $E_\mathrm{Th}$) approximately equals 1.15 for the S
case, while for S$_0$ the curve is closer to unity. In the S$_B$ case, $G_{NS}(V)$ monotonically decreases.}
 \label{fig:G_NS_V}
\end{figure}

\textbf{Transparent limit:} $G_N \ll G_T$. In this case, we can retain only the first term in the r.h.s.\ of Eq.\
(\ref{GNS}).

At $eV \ll E_\mathrm{Th}$ we calculate a small correction to the zero-bias conductance. For S and S$_0$ we obtain
\begin{equation}
\frac{G_{NS} (V)}{G_0} = 1+ A \left( \frac{eV}{E_\mathrm{Th}} \right)^2
\end{equation}
with the positive coefficient
\begin{equation}
A = \frac 4{\theta_S^4} \left( \frac 12 + \frac{\sin^2 \theta_S}3 + \frac{3 \sin 2\theta_S}{4\theta_S} - \frac{2\sin^2
\theta_S}{\theta_S^2} \right).
\end{equation}
$A(\theta_S)$ monotonically grows from zero at $\theta_S = 0$ to approximately $0.015$ at $\theta_S =\pi/2$.

The low-bias conductance for S$_B$ is
\begin{equation}
\frac{G_{NS} (V)}{G_0} = \frac{\vartheta_S}{\tanh\vartheta_S} -B \left( \frac{eV}{E_\mathrm{Th}} \right)^2
\end{equation}
with the positive coefficient
\begin{equation}
B = \frac 2{\vartheta_S^2 \tanh^2 \vartheta_S} + \frac 3{\vartheta_S^3 \tanh\vartheta_S} + \frac{3 \cosh^2
\vartheta_S}{\vartheta_S^4} - \frac{4 \sinh 2\vartheta_S}{\vartheta_S^5}.
\end{equation}
$B(\vartheta_S)$ is a monotonically growing function starting from zero at $\vartheta_S = 0$. Although $\vartheta_S$ is
an unknown parameter, it can in principle be determined from measurements in the tunneling limit [see Eq.\ (\ref{ts})
with $\nu_S(0) = \cosh\vartheta_S$].

At $eV \gg E_\mathrm{Th}$ all the three cases (S, S$_0$, and S$_B$) are treated in a similar manner. Since $G_N \ll
G_T$, the boundary condition (\ref{bc0}) at $x=0$ reduces to $\theta = \theta_S$. The well-known solution of the
sine-Gordon equation (\ref{Usadel}) with fixed surface value is
\begin{equation} \label{4atan}
\theta(x) = 4 \arctan \biggl\{ \tan\left( \frac{\theta_S}4 \right) \exp\biggl( -(1-i) |x| \sqrt{\frac ED} \biggr)
\biggr\}.
\end{equation}
This solution satisfies the boundary condition $\theta(-L) = 0$ with good accuracy, because $\theta(-L)$ is
exponentially small at $E \gg E_\mathrm{Th}$. Substituting Eq.\ (\ref{4atan}) into the first term in the r.h.s.\ of Eq.\
(\ref{GNS}), we obtain
\begin{equation}
\frac{G_{NS}(V)}{G_0} = 1 + \int_{-\infty}^0 \frac{dx}L  \tanh^2 \theta_2 (x) = 1 + C \sqrt{\frac{E_\mathrm{Th}}{eV}},
\end{equation}
where the positive coefficient $C$ depends only on $\theta_S$, i.e., on the type of the superconductor. In the S case,
$C\approx 0.3$.

The results for the transparent limit are summarized in Fig.~\ref{fig:G_NS_V}. Note that in \cite{BVElong} and
\cite{BVE_review} a different, so-called cross geometry was considered under assumption of weak proximity effect. Then a
small correction to the normal-state conductance due to the S$_B$ component was numerically shown to monotonically
decrease as a function of temperature at zero bias.

In conclusion, the conductance of the junction between a normal metal and a Berezinskii superconductor (odd-$\omega$
spin-triplet \textit{s}-wave state) has been studied. The main differences from the case of a conventional
superconductor are: (i)~in the tunneling limit, $G_{NS}(V)$ is larger than the normal-state conductance
(Fig.~\ref{fig:G_NS_V_tun}), and (ii)~in the transparent limit, $G_{NS}(V)$ monotonically decreases
(Fig.~\ref{fig:G_NS_V}). These features can be used as an experimental test for the Berezinskii superconductivity in
bulk samples (e.g., Na$_x$CoO$_2$) or in superconductor-ferromagnet proximity systems.

I am grateful to M.\,V.\ Feigel'man and I.\,I.\ Mazin for drawing my attention to this problem and for helpful
discussions. I am also grateful to A.\,A.\ Golubov, Y.\ Tanaka, and A.\,F.\ Volkov for useful discussions of the
results. The research was supported by the Dynasty Foundation, CRDF, the Russian Ministry of Education, RF Presidential
Grant No.\ MK-4421.2007.2, RFBR Grants Nos.\ 07-02-01300 and 07-02-00310, and the program ``Quantum Macrophysics'' of
the RAS.


\begin{thebibliography}{99}
\bibitem{z}
Of course, the direction in the spin space can be chosen arbitrarily, hence the difference from papers considering the
long-range triplet component (LRTC) in SF structures (where the $z$ axis is usually chosen as the direction of the
exchange field in ferromagnetic regions with constant magnetization, while the LRTC corresponds to $\hat\sigma_1$ or
$\hat\sigma_2$ \cite{BVE_review,FVE}) is only in notations.
\end{thebibliography}
\end{document}